\documentclass[aps,prl,10pt,twocolumn,groupedaddress,notitlepage,showpacs,floatfix,longbibliography]{revtex4-1}
\pdfoutput=1
\usepackage{graphicx,graphics,epsfig,subfigure,times,bm,bbm,amssymb,amsmath,amsfonts,amsthm,mathrsfs,MnSymbol}
\usepackage{gensymb}
\usepackage[matrix,frame,arrow]{xypic}
\usepackage[pdfstartview=FitH]{hyperref}
\usepackage{subfigure}
\usepackage{float}
\usepackage{comment}
\hypersetup{
    colorlinks=true,       
    linkcolor=red,          
   citecolor=magenta,        
    filecolor=magenta,      
    urlcolor=cyan,           
    runcolor=cyan
}
\usepackage[pdftex]{color}
\usepackage{braket}
\usepackage{enumerate}
\usepackage[normalem]{ulem}
\usepackage[usenames,dvipsnames]{xcolor}
\usepackage{multirow}
\usepackage{mathtools}

\usepackage{flushend}
\usepackage{orcidlink}
\begin{document}
\title{Quantum connectivity of quantum networks}

\author{Md Sohel Mondal\,\orcidlink{0009-0006-2915-2309}}\thanks{These authors have contributed equally to this work}
\affiliation{Department of Physics and Center of excellence in Quantum Information, Computation, Science and Technology, Indian Institute of Technology Bombay, Powai, Mumbai - 400076, India}

\author{Shashank Shekhar\,\orcidlink{0009-0008-8926-012X}}\thanks{These authors have contributed equally to this work}
\affiliation{Department of Physics and Center of excellence in Quantum Information, Computation, Science and Technology, Indian Institute of Technology Bombay, Powai, Mumbai - 400076, India}

\author{Siddhartha Santra\,\orcidlink{0000-0001-6695-3929}}
\affiliation{Department of Physics and Center of excellence in Quantum Information, Computation, Science and Technology, Indian Institute of Technology Bombay, Powai, Mumbai - 400076, India}

\begin{abstract}
The practical utility of a quantum network depends on its ability to establish entanglement between arbitrary node pairs with quality sufficient to execute entanglement enabled tasks. This capability can be assessed globally, through aggregate performance over all node pairs, as well as locally, at the level of individual nodes. Since entanglement-based connections form a layer above the underlying physical topology, quantum connectivity is not adequately captured by classical topological connectivity metrics. To enable characterisation of the quantum connectivity at the level of the network (or its subnetworks), we introduce the quantum connectivity measure (QCM), which quantifies the average connection quality between pairs of network nodes. Further, we describe two quantities, the quantum-connected fraction (QCF) and the quantum clustering coefficient (QCC), naturally derived from the QCM, which capture important features of the functional connectivity of the quantum network at the level of the network and an individual node, respectively. These metrics of quantum connectivity depend crucially on the entanglement distribution protocol and the quantum network parameters in addition to its physical topology. We demonstrate the crucial distinction between topological and quantum connectivity, showing that even a fully connected graph can be functionally disconnected for quantum tasks if average network edge-concurrence falls below a critical threshold. These quantum connectivity metrics thus provide important tools for the design, optimization, and benchmarking of future quantum networks.
\end{abstract}
\maketitle

{\it Introduction}: Quantum networks (QN) \cite{Kimble2008Qnintro,Wehner2018Qnintro} are envisioned as futuristic communication networks that harness entanglement between quantum systems to transmit and process information, enabling functionalities beyond the scope of classical networks ranging from unconditionally secure communication \cite{bennettQSCintro,Cao2022QKDintro,Ekert1991QKDintro,BENNETT2014QKDintro} to distributed quantum computing and sensing \cite{Avron2021DQCintro,Cirac1999DQCintro,Cacciapuoti2020DQCintro,caleffi2024distributed,Vittorio2004QESintro}. While recent work has shed some light on the properties of quantum networks that distinguish them from classical networks in terms of their deviation from the small-world property \cite{QN_stat_prop}, capacity phase transitions \cite{QN_cap_trans}, noise-robustness \cite{robustness_noisy-networks} and percolation thresholds \cite{conc_perc_th} - a better understanding of their connectivity structure, useful for quantum information processing tasks \cite{teleportation_fid_indranil}, can be useful towards their design and implementation.

The quantum connectivity between pairs of nodes in a quantum network depends crucially on the entanglement distribution protocol \cite{Azuma2023QRintro} utilised by the network. These protocols employ entanglement manipulations of short-ranged states over the network edges via swapping \cite{Pan1998swap,swapping_sohel} and purification \cite{purification_deutsch} to obtain long-range entangled states between distant network nodes. In fact, access to these operations enable two network nodes to be considered quantum-connected even without a direct edge, provided there exists a path of short-range entangled links that can yield a sufficiently high-quality entangled state between them via some distribution protocol.

Therefore, a quantum network can exhibit two forms of connectivity. The first is the \emph{topological connectivity}, determined purely by the existence of paths between node pairs in the underlying graph. This notion is classical in nature and captures whether any route, regardless of quality, exists between two nodes. The second is the \emph{functional connectivity}, which is another layer on top of the topological connectivity and is determined by the ability of a node pair to perform a quantum information processing (QIP) task using the entanglement available between them. A pair of nodes is said to be functionally connected if the effective entanglement between them is of sufficient quality to meet the requirements of the intended QIP task. Classical network measures such as the clustering coefficient \cite{Watts1998} and giant component fraction \cite{barabasi} characterize connectivity purely at the level of the physical graph, and therefore fail to capture the effective functional connectivity that can arise in quantum networks through entanglement distribution protocols \cite{net_stack_wehner}. 

In this letter, we describe the \emph{quantum connectivity measure (QCM)} - a metric of a network's functional connectivity which captures the average strength of the entanglement connection across all node pairs in a network (or its sub-networks). From this primary measure, we derive two complementary quantities that capture distinct aspects of quantum network connectivity. First, we define the \emph{quantum-connected fraction (QCF)}, which measures the proportion of node pairs whose effective entanglement exceeds the task-specific threshold for successful quantum communication. Second, we introduce a quantum analogue of the clustering coefficient, constructed from the QCM, which quantifies the interconnectivity among the neighbours of a node in terms of functional connectivity. We refer to this quantity as the \emph{quantum clustering coefficient (QCC)} of that node. 

In general, the QCM increases smoothly with the average concurrence of network edge-states ($\bar{c}$), whereas, the QCF exhibits a discontinuous behavior with sharp transitions at certain values of $\bar{c}$ depending on the physical network topology and the entanglement distribution protocol. The QCC of a given network node, of course, depends on the functional connectivity in its neighborhood. We illustrate the evaluation of these measures for a family of statistically defined quantum networks. Further, we show the utility of these measures by analyzing the QCM and QCF of distinct topologies at the physical layer such as the fully connected and random quantum networks. Together, these measures enable a quantitative characterization of functional connectivity in quantum networks that is inaccessible to classical topology-based metrics.

\textit{Family of statistically defined quantum networks}:
A quantum network consists of spatially separated nodes connected by physical links that enable the establishment of bipartite entanglement. 
Each link can be characterized by one or more edge-parameters such as its concurrence, entanglement generation probability, latency etc., relevant for an analysis of the entanglement distribution process in the network. The network itself can then be modeled as a weighted graph $G=(V,E,\mu)$, where $V$ is the set of nodes, $E$ is the set of edges, and $\mu=\{\mu_k\}_{k\in E}$ denotes the collection of edge-parameters. To capture the stochastic nature of entanglement generation, we model each $\mu_k$ as a random variable drawn from a probability distribution $p_\mu$ and for simplicity focus on a single characteristic of a network edge - its concurrence.
While the microscopic configuration ${\mu_k}$ may differ across realizations, the macroscopic behavior of the network is determined by the underlying distribution $p_\mu$. Different realizations of the network with the same topology and identical edge-parameter distribution $p_\mu$ form a family of quantum networks, differing only in the specific permutation of edge-parameters.

{\it Quantum connectivity measure - Characterizing the functional connectivity of a set of network nodes}: Unlike classical networks, where connectivity is determined solely by the presence of topological links, quantum nodes can establish an effective direct connection via some entanglement distribution protocol using the available paths and in a topologically connected quantum network, multiple paths can exist between any pair of nodes \cite{mpep_sohel}. We call the network path generating the strongest entangled connection between the node pair as the \textit{optimal path} between the two end nodes. Given a subset of network nodes, $\mathcal{N} \subseteq V$, and a pair of nodes $i,j \in \mathcal{N}$, let $\mathcal{P}_{ij}$ denote the optimal path connecting $i$ and $j$. The effective \emph{connection strength} ($\mathcal{S}_{ij}$) of the node pair is then defined as $\mathcal{S}_{ij}(\{\mu_{k}\}_{k\in\mathcal{P}_{ij}}) := \mathcal{D}(\{\mu_{k}\}_{k\in\mathcal{P}_{ij}})$, where $\mu_k$ is the edge-parameter of the edge $k$ along the path $\mathcal{P}_{ij}$ and $\mathcal{D}:\mu\mapsto\mathbb{R^+}$ maps the set of edge-parameters to path-parameter according to a given entanglement distribution protocol. For example, if the edge-parameter is considered to be the concurrence $(c)$ of pure entangled states shared along the network edges and the entanglement distribution protocol is entanglement swapping along a network path, the effective connection strength of the node pair becomes the path-concurrence after swapping at the intermediate nodes along the optimal path, i.e., $\mathcal{S}_{ij}=\prod_{k\in\mathcal{P}_{ij}}c_k$. We say that a node pair $(i,j)$ is \emph{functionally connected} if the connection strength of the entanglement between them is of sufficient quality to support a given QIP task, whose required minimum strength is specified by a threshold $0 \le \epsilon \le 1$.

Now we define the QCM of the set $\mathcal{N}$ as the average connection strength taken over all node pairs in $\mathcal{N}$:
\begin{align}
\mathcal{Q}_\mathcal{N}(\{\mathcal{S}_{ij}\}_{i,j \in \mathcal{N}}) := \frac{1}{\mathcal{N}_P} \sum_{i,j \in \mathcal{N}} \mathcal{S}_{ij}~\Theta[\mathcal{S}_{ij}-\epsilon].
\end{align}
Here, $\mathcal{N}_P=|\mathcal{N}|(|\mathcal{N}|-1)/2$ is the total number of node pairs in the set $\mathcal{N}$ and $\Theta[\mathcal{S}_{ij}-\epsilon]$ is Heaviside step function which evaluates to 1 only when the connection strength $\mathcal{S}_{ij}$ along the optimal path exceeds the QIP threshold $\epsilon$ and to 0 otherwise. The Heaviside factor ensures that only functionally connected pairs contribute to the average. The QCM thus quantifies the effective strength of \emph{functional connectivity} within the set $\mathcal{N}$ considering the strongest quantum connection via the optimal path for each node pair which may be distinct from the topologically shortest path between two nodes. Therefore, QCM provides a succinct measure of how well a subset of nodes can be interconnected through the quantum network, capturing both the underlying topological structure and the effectiveness of the entanglement distribution protocol given the network edge-parameters. 


{\it Quantum connected fraction:} As a quantity naturally derived from the QCM one can obtain another the QCF,
\begin{align}
    \mathcal{F}_\mathcal{N}(\{\mathcal{S}_{ij}\}_{i,j \in \mathcal{N}}):=\sum_{i,j\in\mathcal{N}}\frac{\partial \mathcal{Q}_\mathcal{N}}{\partial \mathcal{S}_{ij}}=\frac{1}{\mathcal{N}_P} \sum_{i,j \in \mathcal{N}} \Theta[\mathcal{S}_{ij}-\epsilon],
\end{align}
which is the 1-norm of the gradient of $\mathcal{Q}_\mathcal{N}$ with respect to the connection strengths $\{\mathcal{S}_{ij}\}$, that counts the number of functionally connected node pairs that meet the QIP threshold irrespective of the connection strength. Relative to the QCM which varies smoothly, the QCF shows discontinuous behavior at certain values of the connection strengths in the network reflecting dramatic changes to its functional connectivity (see supplementary material). Together, the QCM and QCF quantify both the average quality and the prevalence of functional connectivity within any chosen subset of the network nodes.

The two measures serve distinct but complementary roles. The QCM ($\mathcal{Q}_{\mathcal{N}}$) allows for a quantitative comparison between different node sets, where a higher value signifies greater efficiency in supporting high-quality quantum tasks. In contrast, the QCF ($\mathcal{F}_{\mathcal{N}}$) quantifies the proportion of node pairs that are functionally connected irrespective of their connection strength beyond $\epsilon$. Note that, QCM and QCF depends on the entanglement distribution protocol and the QIP task threshold $\epsilon$. For a given network topology both QCM and QCF can show different behavior of connectivity for different entanglement distribution protocols and different QIP tasks. Hence, these two metrics will be practically useful for designing appropriate entanglement distribution protocols in a network for the feasibility of target QIP tasks.

The two quantities are normalized to the range $[0, 1]$. They vanish if no node pair in $\mathcal{N}$ meets the QIP threshold (i.e., $\mathcal{S}_{ij} < \epsilon, \forall i,j \in \mathcal{N}$) and approach unity for near-perfect connections. When evaluated over the entire network ($\mathcal{N}=V$), $\mathcal{F}_V$ acts as a quantum analogue of the giant component in classical graph theory. It is important to note the distinction: while a giant component refers to the largest physically (i.e., topologically) connected cluster, the QCF is a global metric that counts all functionally connected node pairs above the threshold, even if they reside in the same, smaller and physically disconnected components of the network \footnote{In classical graph theory the largest connected component in the network is referred to as giant component of the network. The nodes residing within the smaller components of the network are ignored while evaluating the giant component fraction as a connectivity measure. However, in quantum networks, we consider the node pairs residing even within these smaller network components can contribute to QCM and QCF if they share an entangled state of significant quality.}.

Importantly, evaluating the measures $\mathcal{Q}_V$ and $\mathcal{F}_V$ for a network with set of nodes $V$ and set of edges $E$ requires computing the optimal path between all node pairs in the network. When map $\mathcal{D}$ is multiplicative over edge-parameters, the problem reduces to a shortest-path computation on logarithmic edge weights. The total complexity of finding the shortest path between all possible node pairs using Dijkstra's algorithm \cite{dijkstra} is $\mathcal{O}(|V||E|\log |V|)$, after which the measures themselves can be evaluated in $\mathcal{O}(|V|^2)$ time. Since the shortest-path computation dominates the pairwise summation for connected graph, the overall time complexity is $\mathcal{O}(|V||E|\log |V|)$.
\begin{figure}
    \centering
    \includegraphics[width=0.95\linewidth]{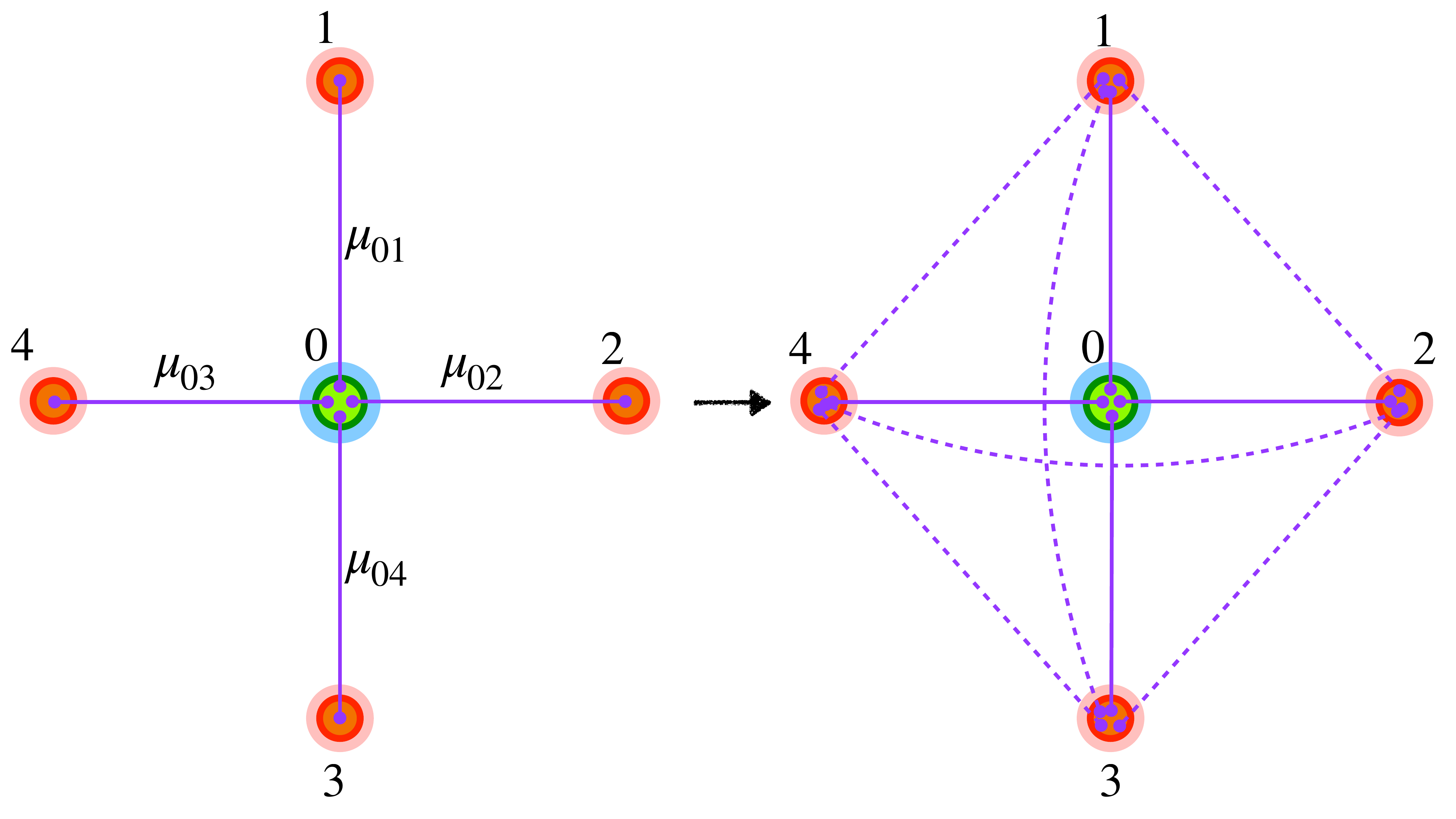}
    \caption{(color online) Illustration of quantum-enhanced connectivity. (Left) A star graph where the neighbors $\{1,2,3,4\}$ of the central node $0$ lack direct connections, resulting in a classical clustering coefficient of zero for node $0$. The edges $(0,i)$ represent quantum channels with edge-parameters $\mu_{0i}$. (Right) By performing entanglement swapping at the central node, effective, all-to-all connections (dashed lines) are established among the neighbors. This creates a fully connected subgraph (a clique), giving rise to a non-zero Quantum Clustering Coefficient.}
    \label{fig:square_network}
\end{figure}

{\it Quantum clustering coefficient of a network node}: Building on the QCM, we can naturally define a \emph{quantum clustering coefficient} (QCC) for a node by considering the connectivity among its neighbors. In classical networks, the clustering coefficient $C_i$ of a node $i$ quantifies how well its neighbors are mutually connected \cite{Holland1971, Watts1998}. In the quantum setting, this mutual connectivity can be evaluated using the QCM of the neighbor set $\mathcal{N}^{(i)}$ of node $i$. Formally, the Quantum Clustering Coefficient of node $i$ is given by,
\begin{align}
\mathcal{C}_Q(i) := \mathcal{Q}_{\mathcal{N}^{(i)}}.
\end{align}
The QCC can take values in the range $[0,1]$ and it vanishes when none of the neighbors of a node can satisfy the QIP task threshold $\epsilon$ pairwise, whereas it becomes unity when all neighbors can be connected via perfect entangled states among themselves. The QCC of a node can reveal significant differences in the connectivity among a node's neighbors compared to classical networks with the same underlying graph topology. 
For example, consider the network shown in Fig.~\ref{fig:square_network}, where the central node $0$ has four neighbors, none of which share direct connections with each other in the given topology. In the classical case, the clustering coefficient of the node $0$ is zero, as its neighbors are completely disconnected. However, in a quantum network with the same topology, effective connections among all neighbors of the node $0$ can be established via entanglement swapping through the node $0$, so that the $\mathcal{C}_Q(0)$ for this node takes a non-zero value. This non-zero $\mathcal{C}_Q(0)$ directly illustrates the advantage of quantum network operations, as it captures potential connectivity among neighbors that is inaccessible in classical networks.

A high $\mathcal{C}_Q(i)$ at a node indicates that its neighboring nodes possess strong quantum connectivity. This property can be exploited when a node ($i$) is untrusted. To ensure security in tasks like quantum secret sharing~\cite{Mark1999QSSintro} and multi-party quantum key distribution~\cite{Epping2017NQKD}, the untrusted node can be bypassed. It can also help in finding efficient routing strategies for quantum communication tasks, such as entanglement-enhanced sensing \cite{Vittorio2004QESintro}, multipartite state generation \cite{multipartite_exp2}, or distributed quantum computation \cite{Avron2021DQCintro}.

{\it Analytical estimation for mean QCM and QCF of network family}: 
The connectivity metrics defined above can be utilized to determine the global connectivity which encapsulates the topological and functional connectivity of a network family given by the probability distribution function (PDF) $p_\mu$ of the edge-parameter. For such a network family the mean QCM can be expressed as:
\begin{align}
\label{eq:general_result_QCM}
  \overline{\mathcal{Q}}^{(G)}
= \int  \mathcal{Q}_V(\{\mathcal{S}_{ij}\}_{i,j \in V})  \prod_{i,j \in V}p_\mathcal{S}(\mathcal{S}_{ij}) d\mathcal{S}_{ij}~,
\end{align}
where $p_\mathcal{S}$ denotes the PDF of the random variable $\mathcal{S}$. For a given network topology with known probability mass function (PMF) $q(\ell_0)$ of the optimal path lengths ($\ell_0$), and the map $\mathcal{D}$ for the entanglement distribution protocol, we can analytically estimate the mean QCM as (see supplementary material),
\begin{align}
\label{eq:iid_result_QCM}
\overline{\mathcal{Q}}^{(G)}= \sum_{\ell_0=1}^{\ell_{\max}} q(\ell_0)
\int_{\mathcal{R}}
\mathcal{D}(\{\mu_{k}\}_{k=1}^{\ell_0})
\left(\prod_{k=1}^{\ell_0} p_\mu(\mu_k)\,d\mu_k\right),
\end{align}
where, we consider independently and identically distributed (i.i.d) random edge-parameters. The integration domain $\mathcal{R}$ is the $\ell_0$-dimensional hypercube $[\text{min}(\mu),\text{max}(\mu)]^{\ell_0}$ constrained by the condition $\mathcal{D}(\{\mu_{k}\}_{k=1}^{\ell_0}) > \epsilon$. This expression averages the connection strength over all edge-parameter configurations that meet the QIP threshold. A parallel derivation yields the mean QCF of the network as,
\begin{align}
\label{eq:general_result_QCF}
    \overline{\mathcal{F}}^{(G)}
= \int  \mathcal{F}_V(\{\mathcal{S}_{ij}\}_{i,j \in V})  \prod_{i,j \in V}p_\mathcal{S}(\mathcal{S}_{ij}) d\mathcal{S}_{ij}~,
\end{align}
The above expression can be reduced to (see supplementary material),
\begin{align}
\label{eq:iid_result_QCF}
\overline{\mathcal{F}}^{(G)}
= \sum_{\ell_0=1}^{\ell_{\max}} q(\ell_0)
\int_{\mathcal{R}}
\left(\prod_{k=1}^{\ell_0} p_\mu(\mu_k)\,d\mu_k\right).
\end{align}
In our analysis we consider quantum networks $G(V,E)$ where each network edge is distributed with a pure bipartite entangled state. The quality of each link is quantified by its \emph{concurrence}. The entanglement distribution protocol considered is the simple entanglement swapping based protocol and hence the concurrence of the resultant entangled state is the product of the individual link concurrences \cite{swapping_sohel}, $\mathcal{D}(\{c_{k}\}_{k=1}^{\ell}) = \prod_{k=1}^{\ell} c_k$ where $c_k$ is the concurrence of the $k^\text{th}$ link. In this setting, Eqs.~(\ref{eq:iid_result_QCM}) and (\ref{eq:iid_result_QCF}) reduces to
\begin{align}
\label{eq:main_result_QCM}
\overline{\mathcal{Q}}^{(G)}
&= \sum_{\ell_0=1}^{\ell_{\max}} q(\ell_0)
\int_{\mathcal{R}}
\left(\prod_{k=1}^{\ell_0} c_k\right)
\left(\prod_{k=1}^{\ell_0} p_C(c_k)\,dc_k\right),\\ 
\label{eq:main_result_QCF}
\overline{\mathcal{F}}^{(G)}
&= \sum_{\ell_0=1}^{\ell_{\max}} q(\ell_0)
\int_{\mathcal{R}}
\left(\prod_{k=1}^{\ell_0} p_C(c_k)\,dc_k\right),
\end{align}
where the integration domain $\mathcal{R}$ is the $\ell_0$-dimensional hypercube $[\min(c),\max(c)]^{\ell_0}$ constrained by the condition $\prod_{k=1}^{\ell_0} c_k > \epsilon$. We consider two kinds of edge-concurrence distributions, one is delta function distribution, $p_C=\delta(c-c_0)$, which corresponds to homogeneous network and another is the uniform distribution, $p_C=U(\min(c),\max(c))$. Here we describe the connectivity of fully connected topology and random network topology for both homogeneous and inhomogeneous distribution of the edge-concurrences.  The expressions of QCM and QCF for fully connected network is analytically obtained in the supplementary material, whereas for random networks the integrations over edge-concurrence involved in Eqs.~(\ref{eq:main_result_QCM}) and (\ref{eq:main_result_QCF}) are performed numerically. 

\begin{figure}
    \centering
    \includegraphics[width=\linewidth]{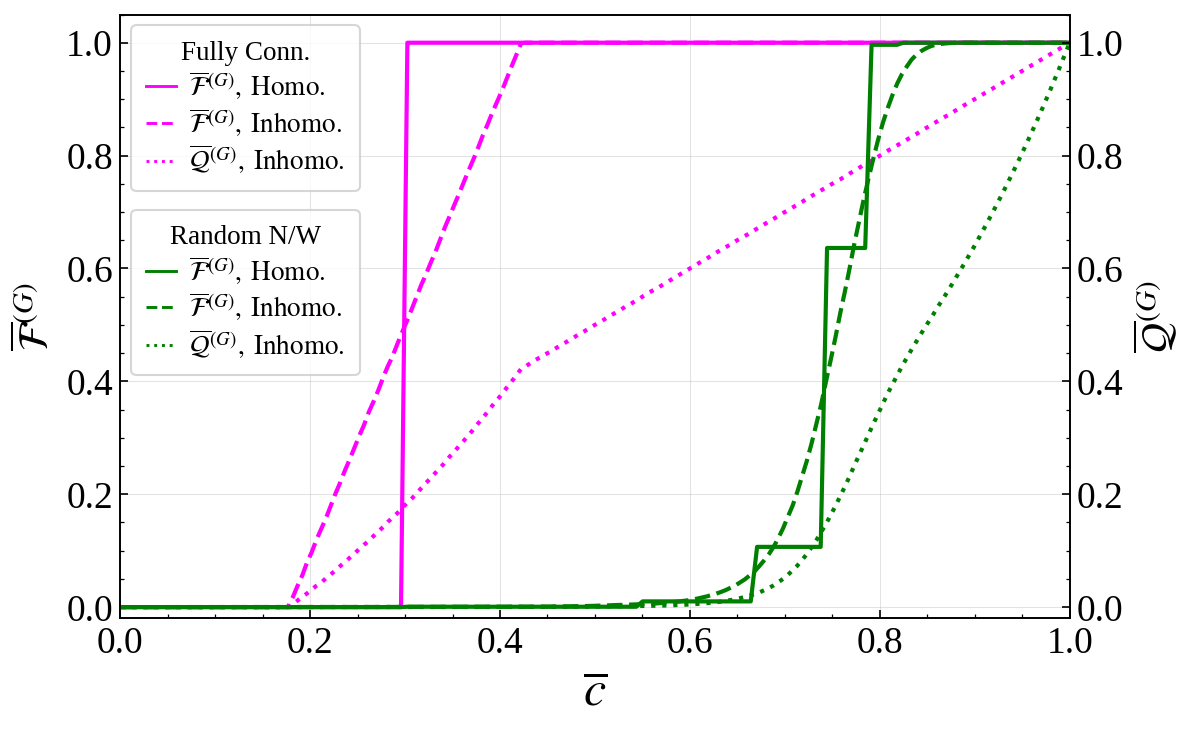}
    \caption{(Color online) Quantum Connectivity Measure (QCM) and Quantum-Connected Fraction (QCF) as a function of the average edge-concurrence $\bar{c}$, for a QIP threshold $\epsilon = 0.3$. Results are shown for two network topologies: fully connected (magenta) and random network (green). The considered random network has $N = 10^4$ nodes and average degree $k = 10$. For each topology, the solid, dashed, and dotted lines represent the QCF for the homogeneous case, the QCF for the inhomogeneous case ($\sigma^2 = 0.005$), and the QCM for the inhomogeneous case ($\sigma^2 = 0.005$), respectively. For the inhomogeneous case, the variance is adjusted near $\bar{c} = 0$ and $\bar{c} = 1$ to ensure that the concurrence distribution remains within $[0, 1]$.}
\label{fig:QCF}
\end{figure}

{\it 1. Quantum connectivity of a topologically fully connected network}: In a topologically fully connected (complete) network, every pair of nodes shares a direct link, so when the variance of the edge-concurrence is small the PMF of the optimal path length can be approximated by a delta function peaked at $\ell_0=1$, i.e., $q(\ell_0)=\delta(\ell_0-1)$. Fig.~\ref{fig:QCF} shows the QCM and QCF for fully connected networks of arbitrary size as a function of the average edge-concurrence $\overline{c}$. The key observation is that, despite being topologically fully connected, the network remains functionally disconnected up to a variance-dependent threshold in $\overline{c}$, demonstrating that classical connectivity does not guarantee feasibility of a given QIP task. Both the QCM and QCF exhibit piecewise behavior depending on the position of the threshold $\epsilon$ relative to the edge-concurrence. Moreover, the QCC of any node in this network equals the QCM of the network since the neighbors of the node again form a fully connected network with the same edge concurrences.

\begin{figure}
    \centering
    \includegraphics[width=0.765\linewidth]{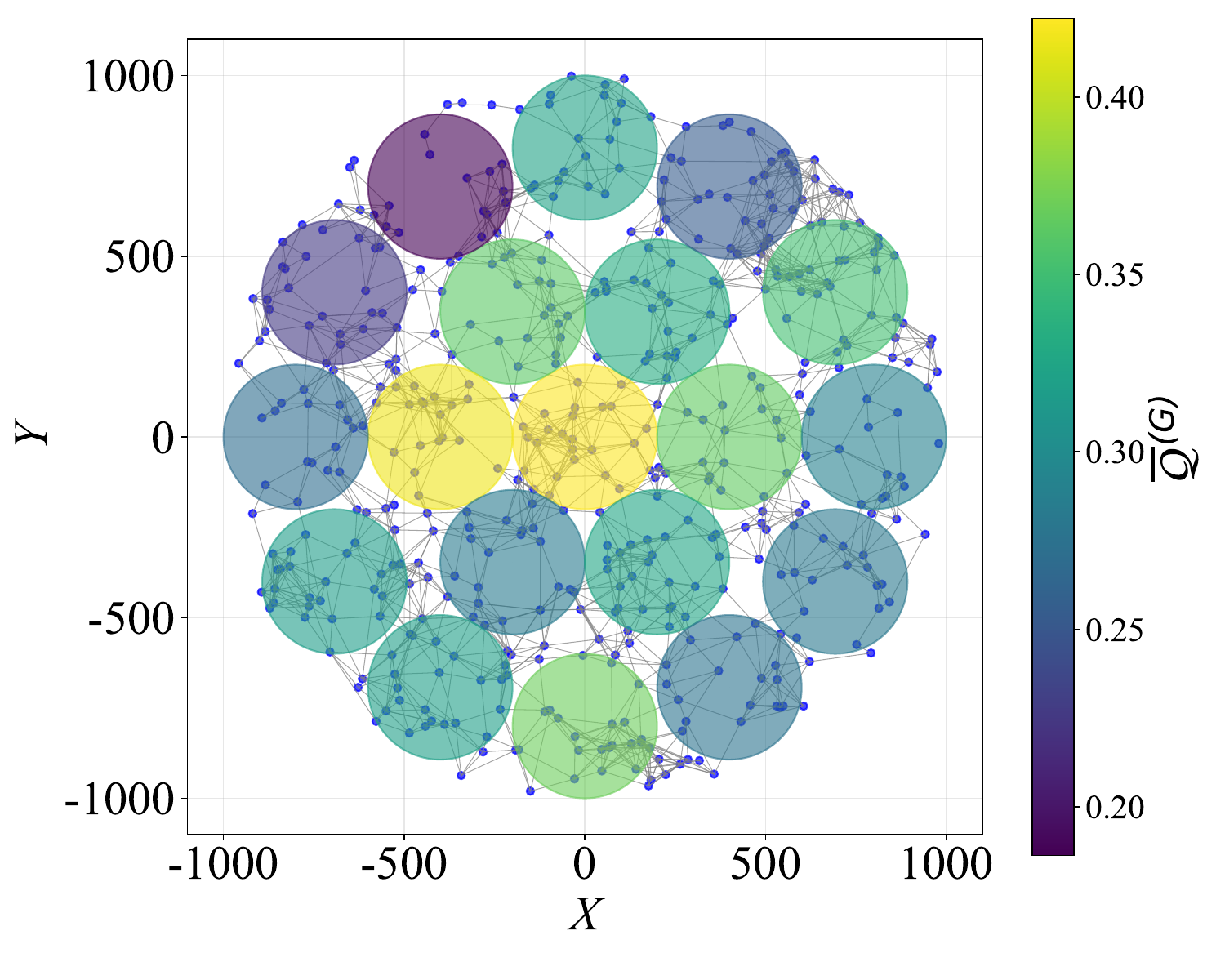}
    \caption{(color online) Waxman network with $N=500$ nodes uniformly distributed within a circular region of radius $1000$ km. Each edge of the network is distributed with an entangled state of concurrence 0.6. The color map indicates the Quantum Connectivity Measure (QCM) for a QIP task with threshold $\epsilon=0.3$, evaluated over smaller partitions of radius $200$ km, highlighting spatial variations in entanglement connectivity across the network.}
    \label{fig:regional_qcm}
\end{figure}

{\it 2. Quantum connectivity of a random network}: We consider a quantum 
network of $N$ nodes with an underlying random graph topology and average 
degree $k$, in the topologically connected phase and therefore every pair of 
nodes is linked by at least one path. For random networks, obtaining an exact PMF for the optimal path is intractable unlike the fully connected case; hence, when the variance of the edge-concurrence is small, we approximate the optimal path to be the shortest graph path (SGP), whose PMF is obtained via network simulations. Fig.~\ref{fig:QCF} shows the QCM and QCF for a random network with $N=10^4$ and $k=10$ as a function of the average edge-concurrence $\overline{c}$. In the homogeneous case, QCF increases in discrete steps as node pairs at progressively larger shortest-path lengths become capable of the task, whereas in the inhomogeneous case the QCF exhibits a smooth transition and saturates at unity after certain value of the average edge-concurrence $\overline{c}$, while the QCM rises continuously and reaches unity only as $\overline{c}\to 1$. The QCC of any particular node in such a random quantum network depends on the functional connectivity of its neighborhood and, in general, varies over the network nodes.

{\it Application of QCM in optical-fiber-based quantum internet}: 
In realistic scenarios, global network connectivity may not always be important; instead, the connectivity of a local region of the network may matter more for many tasks. QCM can be used to study such spatial variations in connectivity across a network. To illustrate this, we consider a photonic quantum network based on an underlying Waxman graph \cite{QN_stat_prop}. In this model, the probability that two nodes separated by Euclidean distance $z$ share a photonic link is $P(z)=e^{-z/2\alpha R}[1-(1-10^{-\gamma z/10})^{n_p}]$, where $\alpha=226/2R$ and $\gamma=0.2~\text{dB/km}$ control the typical edge length and fiber loss, respectively, and $n_p=10^3$ is the number of photons used in an entanglement generation attempt. We consider $N=500$ nodes uniformly distributed over a circular region of radius $R=10^3$ km. The resulting node density $\rho=1.59\times10^{-4}$ per sq. km is above the critical density $\rho_c=6.82\times10^{-5}$ per sq. km, ensuring that the network is connected. Each link is assumed to share a pure entangled state with concurrence $0.6$.

Fig.~\ref{fig:regional_qcm} shows the QCM for a QIP task with threshold $\epsilon=0.3$, obtained by partitioning the network into regions of radius $200$ km. The connectivity of nodes inside a region is not restricted to paths within that region; paths may go through any node in the network. Regions with blue colors have QCM below $0.3$, meaning that node pairs there cannot perform the task on average. Regions with higher QCM can support the task. Yellow regions have the highest QCM and support the task with better quality connections on average, while green regions correspond to the same of a lower-quality.

\textit{Discussions and Conclusion:} The quantum connectivity measures, QCM, QCF and QCC, are numerically and, in some cases, analytically estimable, and allow the quantification of the functional connectivity, that is, the ability of a network's nodes to perform a quantum information processing task given information about the entanglement along the network's edges, beyond classical metrics that characterize physical-layer topology, at both the network level (QCM and QCF) and local levels (QCC). Further, average values of these metrics enabling characterisation of entire families of statistically defined quantum networks can be obtained, as we have shown.

While the QCM varies smoothly with edge-entanglement parameters and increases monotonically with the average concurrence, the QCF exhibits step-like behavior at critical concurrence values which may be useful as targets for achieving a desired fraction of functionally connected node pairs while designing a quantum network. The QCC, in turn, captures functional connectivity in the local neighborhood of a node.

Together these metrics show that the functional connectivity of a quantum network can differ fundamentally from that suggested by its classical topology for a given entanglement distribution protocol and a QIP task threshold. A network that is topologically connected may nevertheless remain functionally disconnected when quantum operations and edge-parameters are taken into account. Conversely, entanglement distribution protocols can induce effective connectivity even when the underlying graph suggests otherwise, making the evaluation of these metrics across different protocols a valuable exercise. Indeed, these metrics can be useful for entanglement distribution protocol design in practical quantum networks for a given target QIP task with known thresholds and potentially can also be utilized for benchmarking network topologies towards implementation of the future quantum internet.

\textit{Acknowledgements}: Funding from DST, Govt. of India through the SERB grant MTR/2022/000389, IITB TRUST Labs grant DO/2023-SBST002-007 and the IITB seed funding is gratefully acknowledged. 


\bibliography{references}

\newpage
\section{Supplementary material}
\subsection{Analytical derivation of quantum-connected fraction.}
\label{app:QCF_derivation}
The quantum connectivity measure (QCM) is expressed as,
\begin{align}
\mathcal{Q}_\mathcal{N}(\{\mathcal{S}_{ij}\}) 
= \frac{2}{|\mathcal{N}|(|\mathcal{N}|-1)} 
\sum_{i,j \in \mathcal{N}} 
\mathcal{S}_{ij}\,\Theta(\mathcal{S}_{ij}-\epsilon),
\end{align}
where $\Theta(x-a)$ denotes the Heaviside function, defined as $\Theta(x-a)=1$ for $x>a$ and $\Theta(x-a)=0$ for $x\le a$. Now we consider the derivative of QCM with respect to the connection strength of a fixed pair $(k,l)$, yielding
\begin{align}
\frac{\partial \mathcal{Q}_\mathcal{N}}{\partial \mathcal{S}_{kl}}
= \frac{2}{|\mathcal{N}|(|\mathcal{N}|-1)} 
\frac{\partial}{\partial \mathcal{S}_{kl}}
\left[\mathcal{S}_{kl}\,\Theta(\mathcal{S}_{kl}-\epsilon)\right].
\end{align}
Using the product rule,
\begin{align}
\frac{\partial}{\partial x}\big(x\,\Theta(x-\epsilon)\big)
= \Theta(x-\epsilon) + x\,\delta(x-\epsilon).
\end{align}
Therefore we obtain,
\begin{align}
\frac{\partial \mathcal{Q}_\mathcal{N}}{\partial \mathcal{S}_{kl}}
= \frac{2}{|\mathcal{N}|(|\mathcal{N}|-1)} 
\left[
\Theta(\mathcal{S}_{kl}-\epsilon)
+ \mathcal{S}_{kl}\,\delta(\mathcal{S}_{kl}-\epsilon)
\right].
\end{align}
Note that, $\delta(\mathcal{S}_{kl}-\epsilon)=0$ for $\mathcal{S}_{kl}\neq \epsilon$. Therefore, for any point except $\mathcal{S}_{kl} = \epsilon$ we can write,
\begin{align}
\mathcal{S}_{kl}\,\delta(\mathcal{S}_{kl}-\epsilon)=0,
\end{align}
and therefore
\begin{align}
\frac{\partial \mathcal{Q}_\mathcal{N}}{\partial \mathcal{S}_{kl}}
= \frac{2}{|\mathcal{N}|(|\mathcal{N}|-1)} 
\Theta(\mathcal{S}_{kl}-\epsilon).
\end{align}

\noindent
Now the gradient of $\mathcal{Q}_\mathcal{N}$ with respect to the connection strengths $\{\mathcal{S}_{ij}\}_{i,j\in\mathcal{N}}$ is given by,
\begin{align}
    \vec{\nabla}_\mathcal{S}\mathcal{Q}_\mathcal{N}=\sum _{i,j\in\mathcal{N}}\frac{\partial \mathcal{Q}_\mathcal{N}}{\partial \mathcal{S}_{ij}} \hat{\mathcal{S}}_{ij},
\end{align}
where, $\hat{\mathcal{S}}_{ij}$ is the unit vector along the direction of $\mathcal{S}_{ij}$ in the hyperspace formed by the connection strengths $\{\mathcal{S}_{ij}\}_{i,j\in\mathcal{N}}$. Considering the 1-norm of the above gradient, we obtain
\begin{align}
\mathcal{F}_\mathcal{N}
:= \sum_{i,j\in\mathcal{N}} 
\frac{\partial \mathcal{Q}_\mathcal{N}}{\partial \mathcal{S}_{ij}}
= \frac{2}{|\mathcal{N}|(|\mathcal{N}|-1)} 
\sum_{i,j \in \mathcal{N}} 
\Theta(\mathcal{S}_{ij}-\epsilon).
\end{align}
The above quantity is called as the QCF which quantifies the total number of node pairs satisfying the QIP threshold in the set $\mathcal{N}$.

\subsection{Expression for mean QCM and QCF of network families}
\label{app:mean QCF and QCM}

We begin with the definition of the mean QCM,
\begin{align}
\overline{\mathcal{Q}}^{(G)}
= \int  \mathcal{Q}_V(\{\mathcal{S}_{ij}\})  
\prod_{i,j \in V}p_\mathcal{S}(\mathcal{S}_{ij}) \, d\mathcal{S}_{ij}.
\end{align}
Substituting the definition of $\mathcal{Q}_V$, we obtain
\begin{align}
\overline{\mathcal{Q}}^{(G)}
= \frac{2}{|V|(|V|-1)} 
\sum_{i,j \in V}
\int \mathcal{S}_{ij}\,\Theta(\mathcal{S}_{ij}-\epsilon)\,
p_\mathcal{S}(\mathcal{S}_{ij})\, d\mathcal{S}_{ij}.
\end{align}

For a given pair $(i,j)$, the connection strength $\mathcal{S}_{ij}$ is determined by an entanglement distribution protocol $\mathcal{D}$ acting on a path of length $\ell_0$ with elementary parameters $\{\mu_k\}_{k=1}^{\ell_0}$, i.e.,
\begin{align}
\mathcal{S}_{ij} = \mathcal{D}(\{\mu_k\}_{k=1}^{\ell_0}).
\end{align}
Assuming i.i.d.\ edge-parameters with PDF $p_\mu$, we rewrite the integral as
\begin{align}\int \mathcal{S}_{ij}\,&\Theta(\mathcal{S}_{ij}-\epsilon)\,
p_\mathcal{S}(\mathcal{S}_{ij})\, d\mathcal{S}_{ij}\nonumber\\
&=\sum_{\ell_0=1}^{\ell_{\max}} q(\ell_0)
\int_{\mathcal{R}}
\mathcal{D}(\{\mu_k\}_{k=1}^{\ell_0})
\prod_{k=1}^{\ell_0} p_\mu(\mu_k)\, d\mu_k,
\end{align}
where $q(\ell_0)$ is the PMF of optimal path lengths and the integration domain
$\mathcal{R}$ is restricted by the condition
$\mathcal{D}(\{\mu_k\}) > \epsilon$ due to the Heaviside function.

Substituting back, we obtain
\begin{align}
\overline{\mathcal{Q}}^{(G)}
= \sum_{\ell_0=1}^{\ell_{\max}} q(\ell_0)
\int_{\mathcal{R}}
\mathcal{D}(\{\mu_k\}_{k=1}^{\ell_0})
\left(\prod_{k=1}^{\ell_0} p_\mu(\mu_k)\,d\mu_k\right).
\end{align}

\vspace{0.5em}

A completely analogous derivation applies to the mean QCF,
\begin{align}
\overline{\mathcal{F}}^{(G)}
= \int  \mathcal{F}_V(\{\mathcal{S}_{ij}\})  
\prod_{i,j \in V}p_\mathcal{S}(\mathcal{S}_{ij}) \, d\mathcal{S}_{ij}.
\end{align}
Substituting $\mathcal{F}_V$,
\begin{align}
\overline{\mathcal{F}}^{(G)}
= \frac{2}{|V|(|V|-1)} 
\sum_{i,j \in V}
\int \Theta(\mathcal{S}_{ij}-\epsilon)\,
p_\mathcal{S}(\mathcal{S}_{ij})\, d\mathcal{S}_{ij}.
\end{align}
Proceeding as before, this reduces to
\begin{align}
\overline{\mathcal{F}}^{(G)}
= \sum_{\ell_0=1}^{\ell_{\max}} q(\ell_0)
\int_{\mathcal{R}}
\left(\prod_{k=1}^{\ell_0} p_\mu(\mu_k)\,d\mu_k\right),
\end{align}
where the domain $\mathcal{R}$ enforces
$\mathcal{D}(\{\mu_k\}) > \epsilon$.

Under the condition when we have entanglement distribution protocol as only entanglement swapping, edge-parameters as concurrence and the states are pure, the concurrence of the end-to-end state is the product of the individual edge-concurrences, so that the connection strength of the optimal path of length $\ell_{0}$ is $\mathcal{S}_{ij} = \prod_{k=1}^{\ell_{0}}c_{k}$ with $c_{k}\stackrel{\rm
i.i.d.}{\sim}p_C(c)$. In this case, the mean QCM and mean QCF takes the form,
\begin{align}
\overline{\mathcal{Q}}^{(G)}
  &= \sum_{\ell_{0}=1}^{\ell_{\max}} q(\ell_{0})
     \int_{\mathcal{R}}
     \left(\prod_{k=1}^{\ell_{0}}c_{k}\right)
     \left(\prod_{k=1}^{\ell_{0}} p_{C}(c_{k})\,dc_{k}\right),
\label{eq:QCM_general}\\
\overline{\mathcal{F}}^{(G)}
  &= \sum_{\ell_{0}=1}^{\ell_{\max}} q(\ell_{0})
     \int_{\mathcal{R}}
     \left(\prod_{k=1}^{\ell_{0}} p_{C}(c_{k})\,dc_{k}\right),
\label{eq:QCF_general}
\end{align}
where the integration domain $\mathcal{R}$ is the
$\ell_{0}$-dimensional hypercube $[\min(c),\max(c)]^{\ell_{0}}$
subject to the constraint $\prod_{k=1}^{\ell_{0}}c_{k}>\epsilon$.
The two expressions differ only in that the QCM integrand carries an
additional factor of $\prod_{k}c_{k}$, the actual connection strength.


\subsection{Quantum connectivity of topologically fully connected network}
In a topologically fully connected network all the network nodes are directly connected to each other and therefore, $q(\ell_0)=\delta(\ell_0-1)$. In this network the average QCM and QCF depends only on the edge-parameter distribution.
For homogeneous distribution of the edge-concurrence, $p_C(c_k)=\delta(c-c_0)$ with average $\overline{c}=c_0$ the QCM and QCF can be written as,
\begin{align}
    \overline{\mathcal{Q}}^{(G)}
  &= \int_{\mathcal{R}}c~\delta(c-c_0)\,dc,\\
\overline{\mathcal{F}}^{(G)}
  &= \int_{\mathcal{R}}\delta(c-c_0)\,dc,
\end{align}
where $\mathcal{R}$ is the region $c>\epsilon$. The value of the above integrations will depend on the value of the edge-concurrence $c_0$. The QCM can be expressed as,
\begin{align}
\overline{\mathcal{Q}}^{(G)} =
\begin{cases}
\overline{c}, & \text{for } \overline{c} > \epsilon \\
0, & \text{for } \overline{c} \leq \epsilon,
\end{cases}
\end{align}
Therefore the QCM is zero when the average edge-concurrence in the network is below the QIP threshold and that increases linearly with the average edge-concurrence when above the threshold. The QCM reaches unity at $\overline{c}=1$. The QCF can be evaluated to be,
\begin{align}
\overline{\mathcal{F}}^{(G)} =
\begin{cases}
1, & \text{for } \overline{c} > \epsilon \\
0, & \text{for } \overline{c} \leq \epsilon.
\end{cases}
\end{align}
The QCF is also zero when the average edge-concurrence is below the QIP threhsold showing similar behavior to the QCM, whereas it suddenly jumps to unity as the average edge-concurrence takes value more than the threshold. Therefore the QCF shows a discontinuous behaviour unlike the QCM.

For inhomogeneous fully connected network with uniform edge-concurrence distribution,
$p_C(c_k)=U(\min(c),\max(c))=\frac{1}{\max(c)-\min(c)}$, the QCM and QCF can be expressed as,
\begin{align}
    \overline{\mathcal{Q}}^{(G)}
  &= \int_{\mathcal{R}}\frac{c}{\max(c)-\min(c)}\,dc,\\
\overline{\mathcal{F}}^{(G)}
  &= \int_{\mathcal{R}}\frac{1}{\max(c)-\min(c)}\,dc,
\end{align}
where $\min(c)=\overline{c}-\sqrt{3\sigma^2}$, $\max(c)=\overline{c}+\sqrt{3\sigma^2}$,
and $\max(c)-\min(c)=2\sqrt{3\sigma^2}$. Depending on the value of the threshold
$\epsilon$ the QCM can be evaluated as,
\begin{equation}
\overline{\mathcal{Q}}^{(G)}=
\begin{cases}
0, & \epsilon>\overline{c}+\sqrt{3\sigma^2},\\[6pt]
\dfrac{\left(\overline{c}+\sqrt{3\sigma^2}\right)^2-\epsilon^2}{4\sqrt{3\sigma^2}},
  & \overline{c}-\sqrt{3\sigma^2}\le\epsilon\le\overline{c}+\sqrt{3\sigma^2},\\[10pt]
\overline{c}, & \epsilon<\overline{c}-\sqrt{3\sigma^2},
\end{cases}
\end{equation}
In this network the QCM is zero when average edge-concurrence is below the threshold, then it increases quadratically with $\overline{c}$ when the QIP threshold resides within the range of the edge-concurrence and the QCM increases linearly reaching unity at $\overline{c}=1$ for the range of edge-concurrence being above the threshold.

The QCF can be expressed as,
\begin{equation}
\overline{\mathcal{F}}^{(G)}=
\begin{cases}
0, & \epsilon>\overline{c}+\sqrt{3\sigma^2},\\[6pt]
\dfrac{\overline{c}+\sqrt{3\sigma^2}-\epsilon}{2\sqrt{3\sigma^2}},
  & \overline{c}-\sqrt{3\sigma^2}\le\epsilon\le\overline{c}+\sqrt{3\sigma^2},\\[10pt]
1, & \epsilon<\overline{c}-\sqrt{3\sigma^2}.
\end{cases}
\end{equation}
The QCF is zero for average edge-concurrence having values below the threshold, then it increases linearly in the region where the threshold is within the range of edge-concurrence and then it becomes unity for the minimum edge-concurrence being above the threshold.

\begin{figure}[H]
    \centering
    \includegraphics[width=\linewidth]{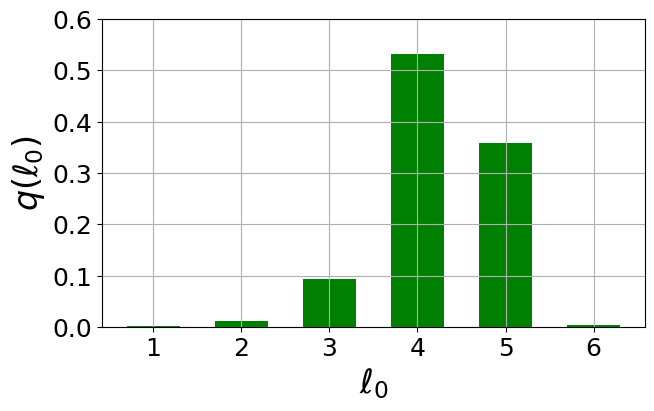}
    \caption{The probability mass function (PMF) of the considered random network with $N=10^4$ nodes and average degree $k=10$.}
    \label{fig:pmf_random}
\end{figure}

\subsection{Quantum connectivity of random network}
For random quantum networks we have approximated that the shortest path is the optimal path to understand the qualitative behavior of QCM and QCF. It is difficult to obtain an analytical expression for the PMF $q(\ell_0)$ of the shortest path length $\ell_0$ . Therefore, for random networks we have obtained the PMF by simulating an actual network and then incorporated the obtained PMF into Eqs. (\ref{eq:QCM_general}) and (\ref{eq:QCF_general}) and the integrations over the edge-concurrence are performed numerically for evaluating the metrics for both homogeneous and inhomogeneous case. In Fig.~\ref{fig:pmf_random} we present the PMF of $\ell_0$ of the random network with number of nodes $N=10^4$ and average degree $k=10$ used in our work.
\vspace*{\fill}
\end{document}